\documentclass[useAMS,usenatbib]{mn2e}

\usepackage{graphicx}
\usepackage{amsmath}
\usepackage{hyperref}
\title[Diagnostics of the Precursor and Main phases]{Multi-wavelength Diagnostics of the Precursor and Main phases of an M1.8 Flare on 2011 April 22}
\author[Arun K. Awasthi et al.]{A. K. Awasthi,$^{1}$\thanks{E-mail: awasthi@prl.res.in (AKA)}
R. Jain,$^{1}$
P. D. Gadhiya,$^{1}$
M. J. Aschwanden,$^{2}$
W. Uddin,$^{3}$
\newauthor
A. K. Srivastava,$^{3}$
R. Chandra,$^{4}$
N. Gopalswamy,$^{5}$
N. Nitta,$^{2}$
S. Yashiro,$^{5}$
\newauthor
P. K. Manoharan,$^{6}$
D. P. Choudhary,$^{7}$
N. C. Joshi,$^{3}$
V. C. Dwivedi,$^{6}$ and K. Mahalakshmi$^{6}$\\
$^{1}$Physical Research Laboratory, Ahmedabad, India, Ahmedabad, India\\
$^{2}$Lockheed Martin Solar and Astrophysics Laboratory, Palo Alto, California, United States\\
$^{3}$Aryabhatta Research Institute of Observational Sciences, Nainital, India\\
$^{4}$Kumaun university, Nainital, India, Nainital, India\\
$^{5}$NASA Goddard Space Flight Center, Greenbelt, Maryland, United States\\
$^{6}$Radio Astronomy Centre, NCRA, TIFR, Ooty, India\\
$^{7}$California State University Northridge, Northridge, California, United States}

\begin{document}

\date{Accepted .............. Received ..............; in original form ............}

\pagerange{\pageref{firstpage}--\pageref{lastpage}} \pubyear{.......}

\maketitle

\label{firstpage}

\begin{abstract}
We study the temporal, spatial and spectral evolution of the M1.8 flare, which occurred in NOAA AR 11195 (S17E31) on 22 April 2011, and explore the underlying physical processes during the precursors and their relation to the main phase. The study of the source morphology using the composite images in 131~$ \rm \AA$ wavelength observed by the {\it SDO/AIA} and 6-14 keV revealed a multi-loop system that destabilized systematically during the precursor and main phases. In contrast, HXR emission (20-50 keV) was absent during the precursor phase, appearing only from the onset of the impulsive phase in the form of foot-points of emitting loop/s. This study has also revealed the heated loop-top prior to the loop emission, although no accompanying foot-point sources were observed during the precursor phase. We estimate the flare plasma parameters viz. temperature (T), emission measure (EM), power-law index ($\gamma$), and photon turn-over energy ($\epsilon_{to}$) and found to be varying in the range of 12.4 - 23.4 MK, 0.0003 - 0.6 $\times$ 10$^{49}$ cm$^{-3}$, 5 to 9 and 14-18 keV, respectively by forward fitting {\it RHESSI} spectral observations. The energy released in the precursor phase was thermal and constituted $\approx$1 per cent of the total energy released during the flare. The study of morphological evolution of the filament in conjunction with synthesized T and EM maps has been carried out which reveals (a) Partial filament eruption prior to the onset of the precursor emission, (b) Heated dense plasma over the polarity inversion line and in the vicinity of the slowly rising filament during the precursor phase. Based on the implications from multi-wavelength observations, we propose a scheme to unify the energy release during the precursor and main phase emissions in which, the precursor phase emission has been originated via conduction front formed due to the partial filament eruption. Next, the heated leftover S-shaped filament has undergone slow rise and heating due to magnetic reconnection and finally erupted to produce emission during the impulsive and gradual phases. 
\end{abstract}


\section{Introduction}

According to the \textquotedblleft standard model\textquotedblright ~of energy release in solar flares, the acceleration of the charged particles takes place following the reconnection of the overlying magnetic field lines. These accelerated electrons lose most of their energy through Coulomb collisions in the chromosphere and emit thick-target hard X-rays (HXRs). As a consequence, the chromosphere is heated up, which enhances the local pressure and thereby drives the heated plasma up into the coronal loops which appear in the form of soft X-rays (SXRs) as a result of thermal bremsstrahlung within the loop plasma. Following this mechanism, non-thermal emission (mostly in HXR energy band) must either accompany or precede the SXR thermal emission \citep{1999Ap&SS.264..129S, 2000BASI...28..117J, 2005SoPh..227...89J, 2011SoPh..270..137J, 2011LRSP....8....6S}. However, \citet{1983BASI...11..318M} showed the appearance of a pre-maximum phase a few minutes before the impulsive phase, termed later as precursor phase and studied by several authors e.g. \citet{1998SoPh..183..339F}, \citet{2002A&A...392..699V}, \citet{2009A&A...498..891B}, \citet{2011ApJ...733...37F}, \citet{2011ASInC...2..297A}, \citet{2012ApJ...758..138A}.\\

Based on multi-wavelength observations, the time evolution of solar flare has been categorized into three phases {\it viz.} precursor, impulsive and gradual. Though, the impulsive and gradual phases (combinely termed as main phase) of solar flares have been studied in greater detail, the underlying processes of energy release in the precursor phase and their relation to the main phase has not been fully established owing to lack of high spatial, spectral and temporal resolution observations during the precursor phase. Further, the physical processes occurring during the precursor phase emission do not form a part of the standard model of energy release in solar flares \citep{1999Ap&SS.264..129S, 2011LRSP....8....6S} and therefore leaves a missing link in the understanding of energy release processes during this phase. The aim of the current investigation is to understand the energy release process in the precursor phase and its association with the main phase.\\

The precursors to the flare are identified based on the disk integrated X-ray emission \citep{1998SoPh..183..339F}. The precursor phase study with spatially resolved observations has revealed interesting insights on the physical processes occurring during this phase and their relation to the main phase. From the spectral analysis of X-ray emission during the precursor phase, \citet{2009A&A...498..891B} found that the precursor emission is of thermal origin and suggested that chromospheric evaporation during this phase is occurring due to the conduction driven saturated heat flux. On the contrary, \citet{2012ApJ...758..138A}, in their study of precursor events have used microwave observations and found the evidences of non-thermal electrons during this phase even when the HXR emission is absent. They suggested microwave observations to be better proxy of non-thermal processes during solar flares than the HXRs. \citet{2011ApJ...733...37F} found that the energy required to produce precursor SXR emission can be completely derived from the energy available with non-thermal electrons while discarding the need of any other heating mechanism. However, their analysis is constrained to single loop model. Therefore different schools of thoughts are being proposed to explain the origin of precursor phase emission. Further, in search of a driver of the instability during the precursor phase, \citet{2006A&A...458..965C, 2007A&A...472..967C} have found slow-rise of the filament that temporally and spatially associated with the precursor brightening (see also \citealt{2011ApJ...743..195J}). Therefore, the origin of precursor phase and its relation to the main phase has been an open issue and a quantitative study of the thermal and non-thermal energetics as well as the filament dynamics employing multi-wavelength observations is very essential.\\

During solar flare energy release, both the thermal and non-thermal emissions are usually observed \citep{2000BASI...28..117J, 2005SoPh..227...89J}. The non-thermal emission is an indication of the particle acceleration, while the thermal emission is a signature of chromospheric heating produced either by non-thermal charge particle beam-driven or by conduction-driven plasma \citep{1985ApJ...289..425F, 1989ApJ...346.1019F, 2012A&A...540A..24B}. \citet{1994PASJ...46L..11N}, in their study of thermal and non-thermal energetics of solar flare using radio and soft X-ray observations found that the heated plasma observed during the impulsive and gradual phases can only be maintained from continuous bombardment of accelerated charge particles for $\sim$ 1000s. \citet{1985BASI...13..253J} conducted the statistical study of 1885 $H \alpha$ flares associated with microwave emission and suggested that two-ribbon (TR) flares and associated impulsive microwave bursts are produced as a consequence of the interaction of non-thermal electrons with ambient plasma embedded in strong magnetic field near lower atmosphere. Using {\it SDO/AIA} observations, \citet{2012A&A...540A..24B} found that chromospheric evaporation is driven via thermal conduction front which is energized by magnetic reconnection. \citet{2005A&A...438..325L} analyzed spectral observations from {\it RHESSI} and found that during the gradual phase, only 20 per cent of the energy transported to the chromosphere has been contributed due to the non-thermally accelerated charged particles, however, the rest came from the charged particles driven through thermal conduction. \citet{2011A&A...535A.123R} performed a correlation study on $H_\alpha$ observation of high cadence with the respective X-ray emission and suggested that longer time-delays between $H \alpha$ and HXR emission of $\approx$20 s correspond to a slow-chromospheric response associated with heating through conduction whereas short delays $\approx$1-2 s are consistent with energy transfer through beam-driven evaporation. Therefore, a correlation study of $H \alpha$ and X-ray emission provides a better opportunity to understand the principle mechanisms of energy transfer in various phases of solar flares. Therefore the spatial, spectral and temporal evolutions of thermal and non-thermal sources is the focus of current study in order to understand the underlying physical processes of energy release in various phases of emission.\\

Here we present the investigation of spatial, spectral and temporal characteristics in various phases of the flare using multi-wavelength observations during the M1.8 flare observed on 2011 April 22. We give a brief description on the observations used for this study in Section 2. Multi-wavelength diagnostics of the precursor and main phases in conjunction with the morphological evolution of the filament is presented in Section 3. We discuss and summarize the study in Section 4.

\section[]{Observations}
The active region (AR) 11195 located at (S17, E31) on 2011 April 22 first appeared at the south-east limb on 2011 April 19. It produced 6 flares of GOES C-class and 2 M-class during 2011 April 20-22. The M1.8 flare that occurred on 2011 April 22 is the subject of current investigation because this event shows very clear precursor enhancement before the impulsive phase as observed in multi-wavelength emissions. In order to investigate the various phases of the flare, we use X-ray observations in 4-6 keV during 04:00:00 UT - 05:10:00 UT from the Si pin detector onboard the Solar X-Ray Spectrometer ({\it SOXS}) mission. {\it SOXS} was launched on board the Indian GSAT-2 spacecraft on 2003 May 8 \citep{2005SoPh..227...89J} and has been successfully operational till 2011 May 2. {\it SOXS} employs Si and CZT solid-state detectors to observe the Sun for 2-3 hours every day. The Si detector provides high spectral resolution ($\sim$0.8 keV) in the 4 - 25 keV energy range. The temporal cadence of observations is 3 s for the quiet activity time which, however, records the X-ray emission with observing cadence of 100 ms during flare activity for 287 s based on the on-board automatic flare detection algorithm after which, the observation cadence is returned to 3 s. The data have been archived on the {\it SOXS} URL http://www.prl.res.in/$\sim$soxs-data. In addition, we use X-ray observations made by the {\it RHESSI} mission in the 6-50 keV energy band from 04:09:00 UT to 05:05:00 UT as the observations were not available during 04:00:00-04:09:00 UT due to {\it RHESSI} night time. The {\it RHESSI} facilitates X-ray imaging as well as spectroscopy of solar flares \citep{2002SoPh..210....3L}. These observations are analyzed with the help of SolarSoftWare (SSW), a software with a repository of codes and graphical user interfaces (GUIs) written in the interactive data language (IDL). Shown in Fig. 1 is the intensity time profile of the flare in 4-6 keV ({\it SOXS}) and in 6-20 as well as in 20-50 keV ({\it RHESSI}) revealing the precursor, impulsive and gradual phases. We note several spikes in the {\it RHESSI} light curve which correspond to the change of attenuators during the observations. The event history is summarized in Table I.\\

In order to explore the spatial correlation of multi-wavelength emission with photospheric magnetic-field observations, we use observations from Helioseismic Magnetic Imager (HMI; \citealt{2012SoPh..275..207S}) onboard the Solar Dynamic Observatory (SDO) mission. It provides full-disk magnetograms at wavelength 6173 ~$ \rm \AA$ with spatial resolution of 0.5'' per pixel and temporal cadence of 45s. We further use EUV observations in 94 and 131 ~$ \rm \AA$ wavelengths observed from Atmospheric Imaging Assembly (AIA; \citealt{2012SoPh..275...17L}) onboard Solar Dynamic Observatory (SDO) combining with the {\it RHESSI} observations to restore the structure of emitting source. The Solar Dynamic Observatory observes the Sun in EUV lines {\it viz.} Fe~{\sc xviii} (94 ~$ \rm \AA$), Fe~{\sc viii, xxi} (131 ~$ \rm \AA$), Fe~{\sc ix} (171 ~$ \rm \AA$), Fe~{\sc xii, xxiv} (193 ~$ \rm \AA$), Fe~{\sc xiv} (211 ~$ \rm \AA$), He~{\sc ii} (304 ~$ \rm \AA$), and Fe~{\sc xvi} (335 ~$ \rm \AA$).The spatial and temporal resolution of these observations are 0.6'' per pixel and 12 s, respectively. We investigate the spatial evolution of temperature (T) and emission measure (EM) synthesized from EUV observations taken by {\it SDO/AIA} in 6 wavelength channels viz. 94, 131, 171, 193, 211, 335 ~$ \rm \AA$ which cover the temperature range of 0.6 to $\sim$16 MK. \\

We study the temporal evolution of $H \alpha$ emission and spatial evolution of filament using observations from Solar Tower Telescope installed at Aryabhatta Research Institute of Observational Sciences (ARIES), Nainital, India \citep{2006JApA...27..267U}. The 15-cm, f/15 $H \alpha$ telescope observes active regions on the Sun with superb time cadence (up to 2s). The images are recorded by a 16 bit 1K$\times$1K pixel CCD camera system having a pixel size of 13 $\rm \mu$m with a spatial resolution of 0.58'' per pixel. This provides unique opportunity to investigate the correlation of temporal evolution of $H \alpha$ with multi-wavelength intensity in various phases of flare. \\

\begin{center}
\begin{table*}
\centering
 \begin{minipage}{\textwidth}
 \centering
  \caption{Time line of activities during the precursor and main phases of the flare.}
  \begin{tabular}{@{}lll@{}}

\hline
S.N & Time (UT)	& Observations/Events\\
\hline
1  & 	04:12:34 - 04:13:50	& Eruption of the South-West leg (L1) of the filament\\
2  &	04:15:18	        & Onset of the precursor phase in 4-6 and 6-14 keV emission\\
3  &	04:18:26	 & First SXR peak during the precursor phase ($P_{x1}$)\\
4  &	04:20:40	 & First $H \alpha$ peak during the precursor phase ($P_{h1}$)\\
5  &	04:27:14 - 04:36:00	& Heating in the vicinity of the filament and enhanced ambient density\\
6  &	04:35:00	& Onset of the filament eruption\\
7  &	04:36:00	& Onset of impulsive phase in 20-50 keV (HXR) emission\\
8  &	04:39:16/04:39:26	& First Peak in HXR and RI in $H \alpha$ time profile after onset \\
 & & of Impulsive phase represented in Fig. 1 by $P_{x2}$ and $P_{h2}$, respectively\\
9  &	04:45:00	& $H \alpha$ plage brightening in the south-east of the AR 11195\\
10 & 	04:54:20	& Commencement of the Gradual Phase\\
11 &	04:54:30	& Peak emission in $H\alpha$ as denoted by $P_{h3}$ in Fig. 1\\
12 &	05:14:00	& End of the Gradual/Decay phase\\
\hline
\end{tabular}
\end{minipage}
\end{table*}
\end{center}

\begin{figure*}
  \centering
  \includegraphics[width=0.9\textwidth, height=10cm]{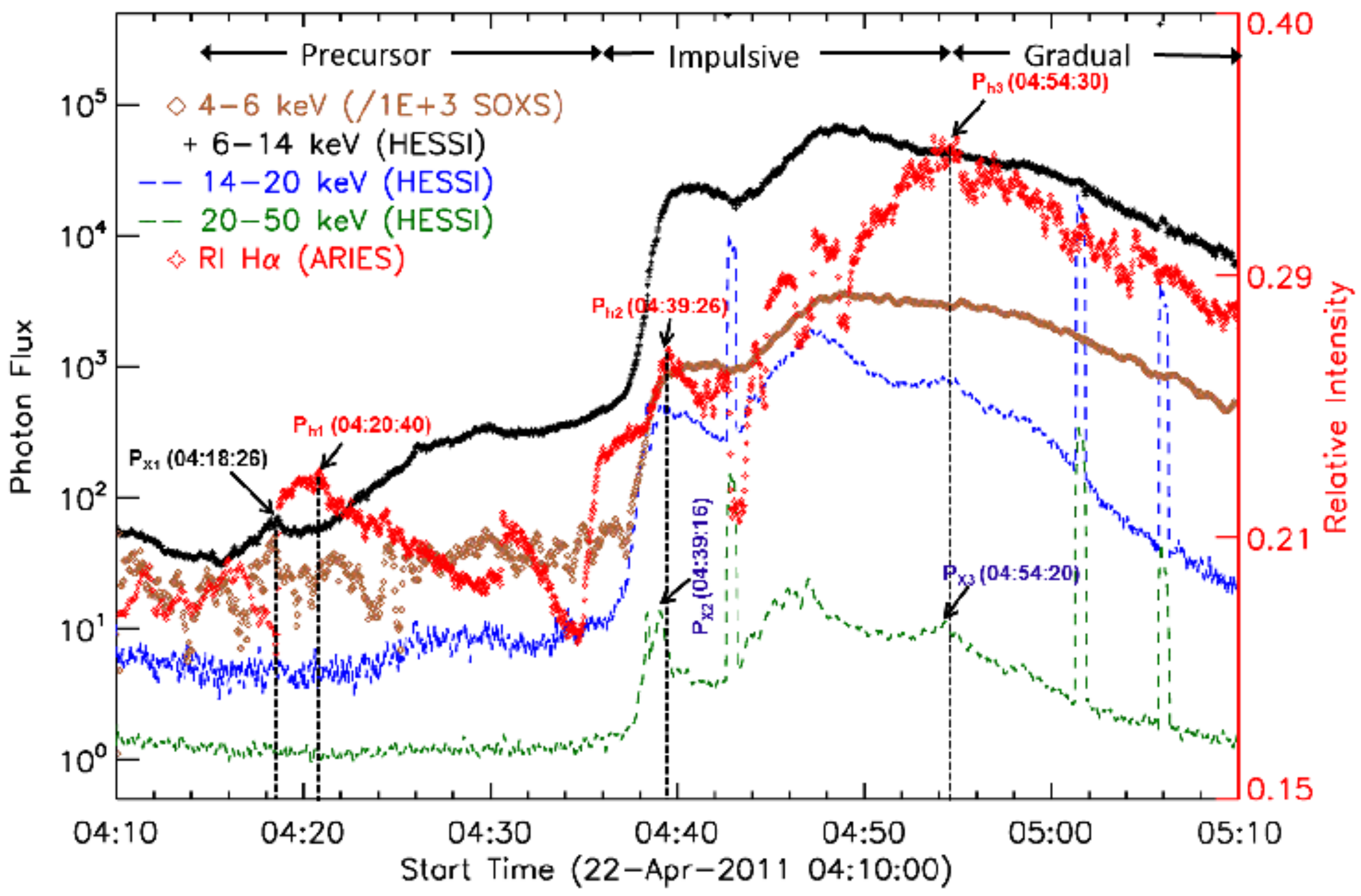}
  \caption{Temporal evolution of multi-wavelength emission during M1.8 flare on 2011 April 22. Relative intensity estimated from $H \alpha$ observations from ARIES/NAINITAL is plotted by red color's symbol. 4-6 keV emission observed by {\it SOXS} as well as 6-14, 14-20 and 20-50 keV emission by {\it RHESSI} shown by brown, black, blue and green symbols, respectively. P$_{x1,2,3}$ and P$_{h1,2,3}$ represent the corresponding peaks during precursor, impulsive and gradual phases in X-ray and $H \alpha$, respectively.}
\end{figure*}

\section[]{Evolution of Multi-wavelength emission in the precursor and main phases}
The precursor phase is identified by the enhanced X-ray emission in SXR ($\textless$20 keV) 10-20 minute prior to the onset of the impulsive phase. Fig. 1 shows temporal evolution of multi-wavelength emission during M1.8 flare on 2011 April 22. 4-6 keV emission observed by {\it SOXS} as well as 6-14, 14-20 and 20-100 keV emissions observed by {\it RHESSI} are shown by brown, black blue and green symbols, respectively. Relative intensity of the $H \alpha$ emission observed by $H \alpha$ telescope at ARIES/NAINITAL is estimated from the following equation.

\begin{center}
$RI$ = ($I_f$ - $I_b$)/$I_b$
\end{center}

Here $I_b$ represents the background intensity estimated from a rectangle of 10 $\times$ 10 pixel away from the flare location and $I_f$ represents mean of the intensities of the region in the frame having intensities above $I_b$ for respective frames. Temporal evolution of $RI$ of $H \alpha$ emission during the flare event is plotted by red color's symbol in fig. 1. P$_{x1,2,3}$ and P$_{h1,2,3}$ represent the corresponding peaks during precursor, impulsive and gradual phases in X-ray and $H \alpha$ intensity profile, respectively. \\

From Fig. 1, we may note that the emission in 4-6 and 6-14 keV energy band (from {\it SOXS} and {\it RHESSI} observations respectively) has commenced on 04:15:18 UT and kept on increasing till 04:36:00 UT in the several steps. In contrast, the HXR ($\textgreater$20 keV) emission has been absent during the precursor phase while commenced at 04:36:00 UT in the form of impulsive burst almost simultaneous to impulsive SXR emission. Thus, we consider impulsive phase onset at 04:36:00 UT that lasted till 04:54:30, after which, gradual phase has commenced. The absence of the HXR emission in the precursor phase of this flare motivates us to probe the origin of the SXR emission. To understand this question, we study the spatial, spectral and temporal evolution of multi-thermal emission in various phases of the M1.8 flare occurred on April 22, 2011.

\subsection[short]{Spatial analysis}

In order to study the spatial evolution of X-ray sources during the precursor and main phases, we synthesize X-ray images of the active region during the flare interval 04.08:30 - 05:05:30 UT on April 22, 2011 using {\it RHESSI} observations. {\it RHESSI} observes X-ray emission from the full solar disk in a wide energy range (3 keV - 17 MeV) with high temporal and energy resolution as well as with high signal sensitivity. Such observations enable us to synthesize the 2D images and spectra in the X-ray band, which provide valuable data for investigation of topological evolution of the thermal and non-thermal sources during flares. We employ CLEAN algorithm \citep{2002SoPh..210...61H} to synthesize the X-ray images during the precursor and main phases. Using the sub-collimators 3F, 4F, 5F, 6F, and 8F, the images are synthesized with spatial resolution of 1'' and integrated over 60 s duration in 4 energy bands viz. 6-10, 10-14, 14-20 and 20-50 keV. The purpose of synthesizing the images in the aforesaid energy bands is to study the morphological evolution of sources of thermal and non-thermal emission. Generally, 6-10 and 10-14 keV emissions are dominated by thermal processes. Further, 14-20 keV emission may be considered to be an intermediate energy band between thermal or non-thermal processes and may be dominated by one component on the other over the evolution of the flare \citep{2005SoPh..227...89J, 2011SoPh..270..137J, 2007ApJ...661.1242A}. However, $\textgreater$20 keV emission is considered to be originated mainly from the population of accelerated electrons and represent non-thermal source morphology. Further, in order to clearly visualize the source topology, we complement the synthesized X-ray images with the observations of {\it SDO/AIA} 131 ~$ \rm \AA$. The temperature response of the 131 ~$ \rm \AA$ channel peaks $\sim$10 MK \citep{2013SoPh..tmp..158N} and therefore represents very hot plasma produced during the flare heating due to thermal processes which are also the origin of SXR ($\textless$10 keV) emission. Fig. 2 shows a sequence of images in 131~$ \rm \AA$ wavelength observed by SDO overlaid by the contours of 30, 50 and 80 per cent intensity (drawn by red, blue and yellow lines respectively) of the maximum emission in co-temporal 6-10 keV energy band during the precursor phase and main phases. Panel (a) of fig. 2 shows 131 ~$ \rm \AA$ images at 04:09:09, 04:12:09 and 04:14:45 UT which represent the topology of AR and loops prior to the precursor phase commencement. We may note from panel (a) that the loop-top appears to be brightened which may refer to the onset of heating at the coronal location of the loop and energizes the transfer of coronal plasma towards the chromosphere through conduction. Panel (b) shows the precursor phase emission evolution at 04:18:33, 04:24:57 and 04:27:09 UT, respectively. Panel (c) shows the images at 04:31:09, 04:35:09 and 04:37:09 UT representing the topological changes during precursor to main phase emission changeover and panel (d) represents the spatial evolution of sources during main phase at 04:40:59, 04:47:00 and 04:55:00 UT. From the panel (b), we may note a uniformly brightened loop $\textquoteleft$L1' with a cusp shaped structure. Further, from panel (c) and (d), it may be noted that as the impulsive phase has commenced, the loop which is brightened during the precursor phase has faded and another loop $\textquoteleft$L2' located close to northern foot-point of $\textquoteleft$L1' started emission. From panel (d), we also note the systematic movement of X-ray emission as well as loops towards the north-west side. This kind of systematic motions in SXR source appears to be associated with the phenomena of asymmetric eruption as already reported for an X2.6 event on 2005 January 15 by \citet{2010ApJ...721L.193L}. \\

\begin{figure*}
  \centering
  \includegraphics[width=0.9\textwidth, height=0.85\textheight]{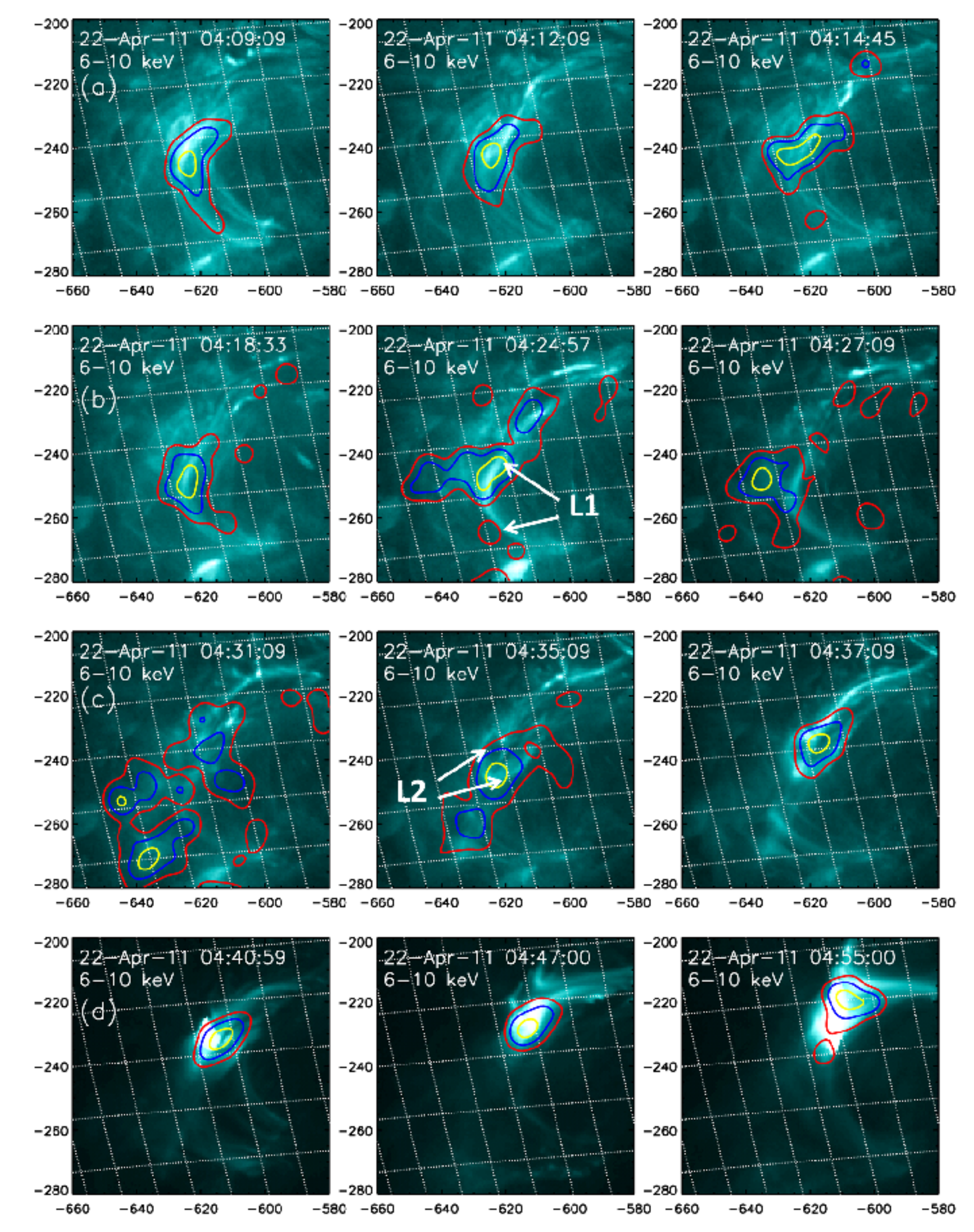}
  \caption{{\it RHESSI} X-ray contours of 30, 50 and 80 per cent intensity of the maximum of 6-10 keV energy band in each frame over plotted on the {\it SDO/AIA} observations in 131~$ \rm \AA$ wavelength. The three contours levels are drawn by red, blue and yellow colors' lines respectively. Panel (a) and (b) represent the evolution of thermal emission during the precursor phase while (c) and (d) represent the same during precursor to main-phase transition and main phase, respectively. The flaring loop during the precursor and impulsive phases is denoted by $\textquoteleft$L1' in panel (b) and by $\textquoteleft$L2' in the panel (c), respectively.}
\end{figure*}

Further, we present the evolution of 10-14, 14-20 and 20-50 keV (HXR) emissions during the precursor and main phase in fig. 3. It is noteworthy that the X-ray intensity profile (cf. fig. 1) did not show the emission in HXR ($\textgreater$20 keV) energy band during the precursor phase which, however, has commenced at 04:36:00 UT, the onset of the impulsive phase. Panel (a), (b), (c) and (d) of the fig. 3 represent the {\it SDO/AIA} images in 131~$ \rm \AA$ wavelength at 04:25:45, 04:38:33, 04:47:00 and 04:55:49 UT, respectively overlaid by contours of 30, 50 and 80 per cent of the intensities in 10-14, 14-20 and 20-50 keV drawn by red, blue and yellow color's lines, respectively. It may be noted from panel (a) of the fig. 3 representing the images during the precursor phase at 04:23:45 UT that we do not see any signature of foot-point emission. However, during the impulsive phase, as shown in panel (b) and (c), we note the foot-points as marked by $\textquoteleft$f1' and $\textquoteleft$f2'. Further, during the evolution of emission in gradual phase, as shown in panel (d), the foot-point separation has increased in comparison to that during impulsive phase which may suggest the rising of reconnection region as modeled in \citet{2001SoPh..204...69F}. Further, the source morphology in both 14-20 and 20-50 keV during the impulsive phase (panel b \& c) reveal foot-point like structure which suggests that 14-20 keV energy band is dominated by non-thermal emission during this phase of the flare.

\begin{figure*}
  \centering
  \includegraphics[width=0.9\textwidth, height=0.85\textheight]{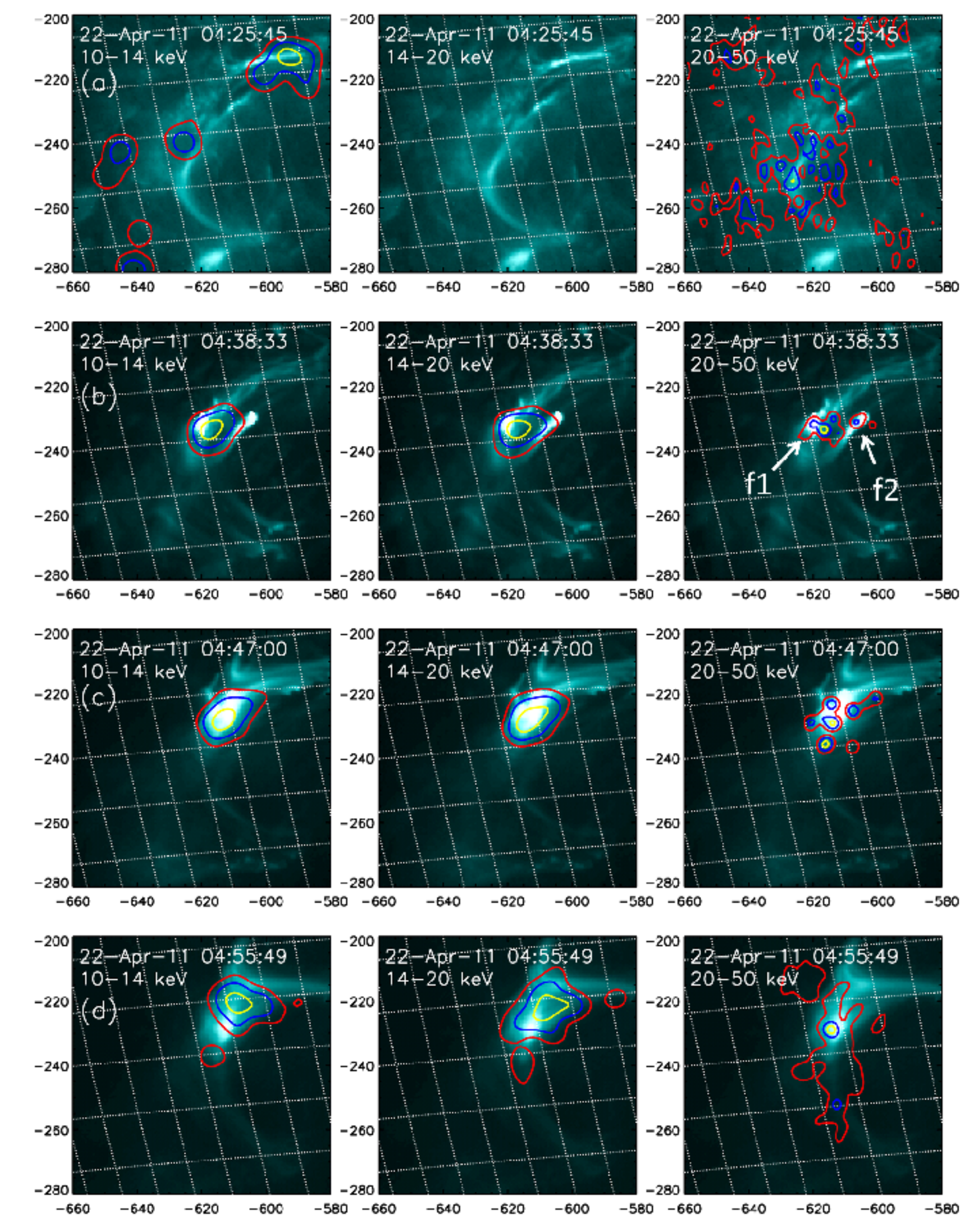}
  \caption{{\it RHESSI} X-ray contours of 30, 50 and 80 per cent intensity of the maximum of 10-14 and 14-20 and 20-50 keV energy bands during the precursor and main phase over plotted on the 131~$ \rm \AA$ emission observations from {\it SDO/AIA} as shown from top to bottom panels respectively. The contours levels are drawn by red, blue and yellow blue colors' lines, respectively. The foot-points revealed by the X-ray images are denoted by $\textquoteleft$f1' and $\textquoteleft$f2'.}
\end{figure*}

\subsection{Spectral and temporal evolution of the X-ray emission}
We study the temporal evolution of the flare plasma parameters viz. temperature (T), emission measure (EM) and density (n$_e$) employing the {\it RHESSI} observations. First, we synthesize the spectrum and spectral response matrix (SRM) files from the X-ray observations during 04:15:00 to 05:10:00 UT of 1 m and 32 s time cadence during the precursor and main phases, respectively. The energy resolution is kept to be 0.3 and 1 keV over the energy band of 6-15 and 15-50 keV, respectively. This spectrum and SRM files serve as input to the object spectral executive (OSPEX\footnote{\url{http://hesperia.gsfc.nasa.gov/ssw/packages/spex/doc/}\\ \url{ospex_explanation.htm}}) package provided in SSW. The OSPEX is an object-oriented interface for X-ray spectral analysis of solar data. In each time-bin, we used OSPEX to estimate the T, EM, power-law index by forward-fitting the observed count spectra with the model count spectra generated by the inbuilt model functions viz. thermal, line emission, multi-thermal, and non-thermal functions which use CHIANTI database \citep{1997A&AS..125..149D, 2012ApJ...744...99L}. The SXR emission is originated either from isothermal or multi-thermal plasma, while HXR emission is considered to be originated via non-thermally excited electrons (during reconnection). As there is no clear-cut demarcated energy between thermal and non-thermal emission, it is estimated by forward fitting the observed spectra fitting simultaneously employing thermal and non-thermal model photon functions. The spectral fitting is carried out by iterative adjustment of the free parameters leading to $\chi^2$ close to 1, which enable us to estimate the flare plasma parameters from the best fitting of modeled counts over the observed counts. \\

\begin{figure*}
  \centering
  \includegraphics[width=0.95\textwidth, height=0.45\textheight]{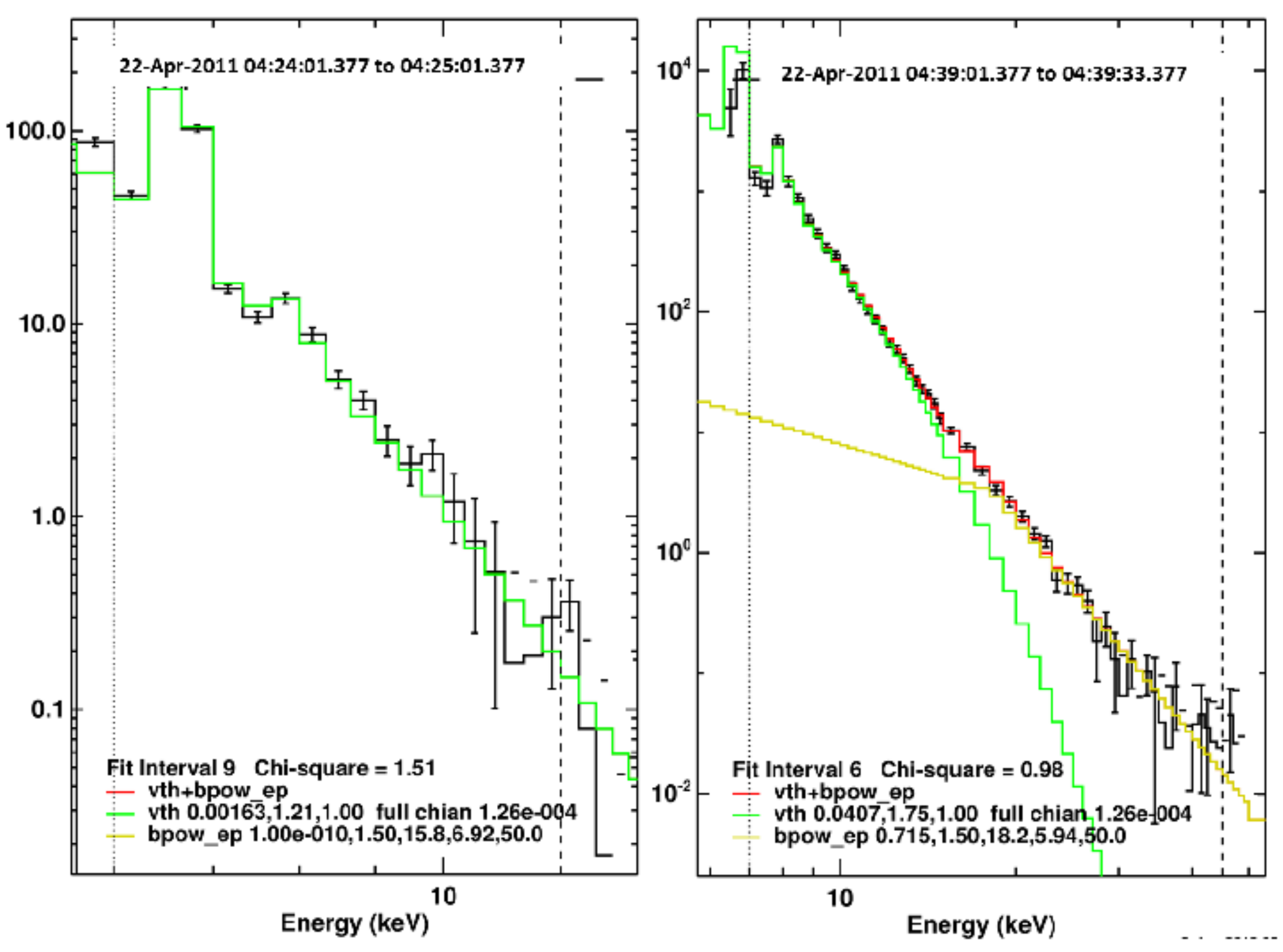}
  \caption{Photon flux spectrum of 2011 April 22 flare observed during 04:24:01-04-25:01 UT and 04:39:01-04:39:33 UT in the left and right panels, respectively shown by the black line. The spectral fit in the left panel is performed employing isothermal photon model function (green line) and that in the right panel employing isothermal (green line) + broken-power-law (yellow line) photon models with goodness of fit represented by $\chi^2$ = 1.51 and 0.98, respectively. The total model photon flux spectrum in the right panel is shown by the red line.}
\end{figure*}

We obtained best fit of the observed flux during the precursor phase considering isothermal function. Left panel of Fig.4 shows the spectral fitting of the X-ray emission in the 6-15 keV energy band (shown by dotted lines) observed during 04:24:01 - 04:25:01 UT, the time of precursor emission. The observed photon flux spectrum (black line) is fitted with the isothermal model (green line) with the goodness of fit represented by $\chi^2$=1.51. The estimated temperature and emission measures are 1.21 keV ($\approx$14 MK) and 0.0016 $\times$ 10$^{49}$ cm$^{-3}$, respectively. The aforesaid technique is applied to the 1 minute and 32 s integrated spectra during 04:15 - 04:36 UT and 04:36 - 05:10 UT, respectively covering all phases of the flare. Thus, we estimated the temporal evolution of the temperature (T) and emission measure (EM) as shown in Fig. 6 panels (a) and (b), respectively. We also show the evolution of density in the panel (b) of fig. 6. The density is estimated by employing the relation

\begin{center}
\begin{math}
n_e=\sqrt{EM/V}
\end{math}
\end{center}

where EM is the emission measure and V is the volume of flaring region. The temporal evolution of EM has already been estimated from the forward-fitting performed on the observed spectra. As the loop morphology is very clear throughout the precursor and main phases, we estimate the volume using cylindrical shape of the loops by $\pi r^2l$ where r is the radius and l is the length of the flaring loop. The length of the loops have been estimated by manually tracing the same from the composite of the observation in 131 ~$ \rm \AA$ wavelength overlaid by contour of 30 per cent of the maximum intensity of co-temporal 6-10 keV energy band and of 94 ~$ \rm \AA$ wavelength drawn by red and blue colors, respectively. The traced loop has been shown by light orange color. The diameter of the traced loop is also calculated from these images by averaging the diameter estimated at three different positions of the visible loop as drawn by yellow line in fig. 5. The first-two images of the Fig. 5 represent the shape of the loop during the evolution of emission in precursor phase at 04:18:21 and 04:26:09 UT. The rest of the two images represent the loop topology during impulsive phase at 04:36:09 and gradual phase at 04:56:12 UT respectively.\\

\begin{figure*}
  \centering
  \includegraphics[width=0.91\textwidth, height=0.16\textheight]{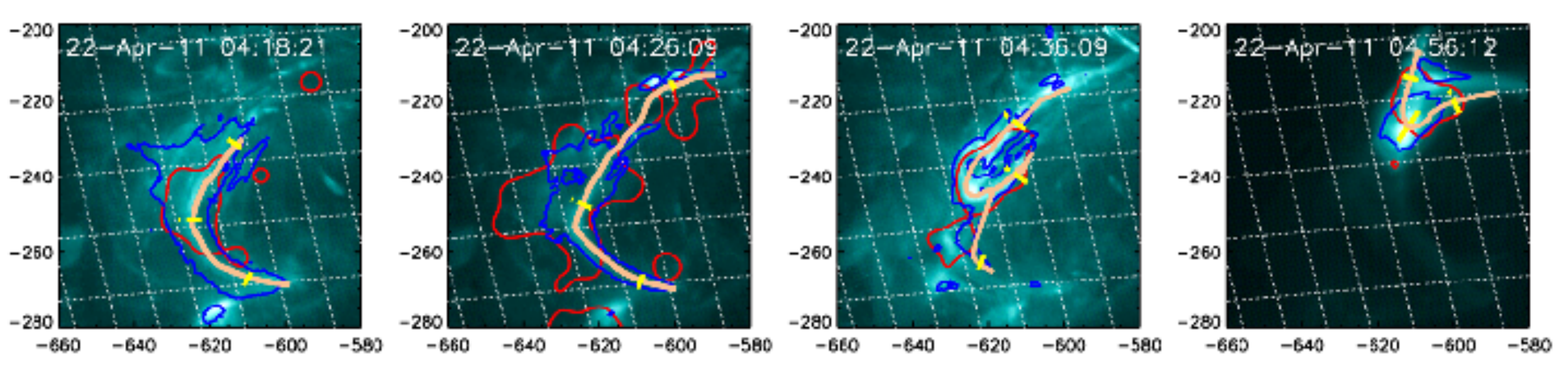}
  \caption{A sequence of images in 131 ~$ \rm \AA$ overlaid by contours of 30 per cent of the maximum intensity of co-temporal 6-10 keV energy band as well as in 94 ~$ \rm \AA$ wavelength drawn by red and blue colors, respectively. Traced loop is drawn by light orange color line while the diameter of the loop is shown by yellow line.}
\end{figure*}

The non-thermal component in spectral observations during the impulsive phase is fitted considering the broken-power law photon model function as shown by the yellow line in the right panel of the Fig. 4. The broken-power law function provides us the photon turn-over energy ($\epsilon_{to}$), power-law index ($\gamma$) above and below the $\epsilon_{to}$ and normalization value at 50 keV ($F_{50}$). The power-law index below $\epsilon_{to}$ is fixed to the value of 1.5 \citep{2005A&A...435..743S}. The temporal evolution of the non-thermal power-law index ($\gamma$) and the HXR flux in 20-50 keV is plotted in the panel (c) and (d) of Fig. 6, respectively. We find T, EM and $\gamma$ to be varying in the range of 12.4 - 23.4 MK, 0.0003 - 0.6 $\times$ 10$^{49}$cm$^{-3}$ and 5 - 9.0, respectively. Further, the photon turn-over energy ($\epsilon_{to}$) is found to be varying in the range of 14-18 keV which is consistent with the similar X-ray source morphologies in 14-20 and 20-50 keV energy bands as presented in section 3.1. From the temporal evolution of non-thermal power-law index ($\gamma$) (cf. panel (d) of Fig. 6), we may note that $\gamma$ remains steady during impulsive phase (04:47:10-04:54:30 UT) to the value of $\sim$7.0. This suggests continuous bombardment of non-thermally accelerated charged particles which may be noted from the panel (d) of the fig. 6. We may further note the double peaks in the HXR intensity profile at 04:39:16UT and 04:47:10 UT followed by a plateau region during 04:47:10-04:54:30 UT. This suggests the involvement of multi-loop configuration followed by their systematic destabilization and activation which is consistent with the movement of the centroids of the X-ray as well as EUV 131~$ \rm \AA$ sources as revealed from the spatial study of multi-wavelength sources in Section 3.1. \\

Further, this study has enabled us to perform a comparative study of the evolution of flare plasma parameters during the precursor and the main phase. The peak temperature during the precursor phase has been estimated to be 19.1 MK, $\sim$80 per cent of that attained during the impulsive phase ($\sim$23.0 MK). It may be noted that the rate of increase of EM during the precursor phase has been insignificant in comparison to that during the impulsive phase. On the other hand, the estimated density has remained almost constant to the value of $\sim$10$^{10}$ cm$^{-3}$ during the precursor phase and increased rapidly after the commencement of impulsive phase to the value $\textgreater10^{11} cm^{-3}$. The above analysis shows slow evolution of T and $n_e$ suggesting gradual heating during the precursor phase which is in agreement to \citet{2009A&A...498..891B}. This may also explain the absence of HXR foot-point emission. We further note from the temporal evolution of $n_e$ as plotted in panel (b) of Fig. 6, it again stabilizes just after the commencement of the gradual phase very similar to that during the precursor phase. This suggests the possibility of common physical origin of the precursor and gradual phases dominated by thermal processes.

\begin{figure*}
  \centering
  \includegraphics[width=0.9\textwidth, height=0.9\textheight]{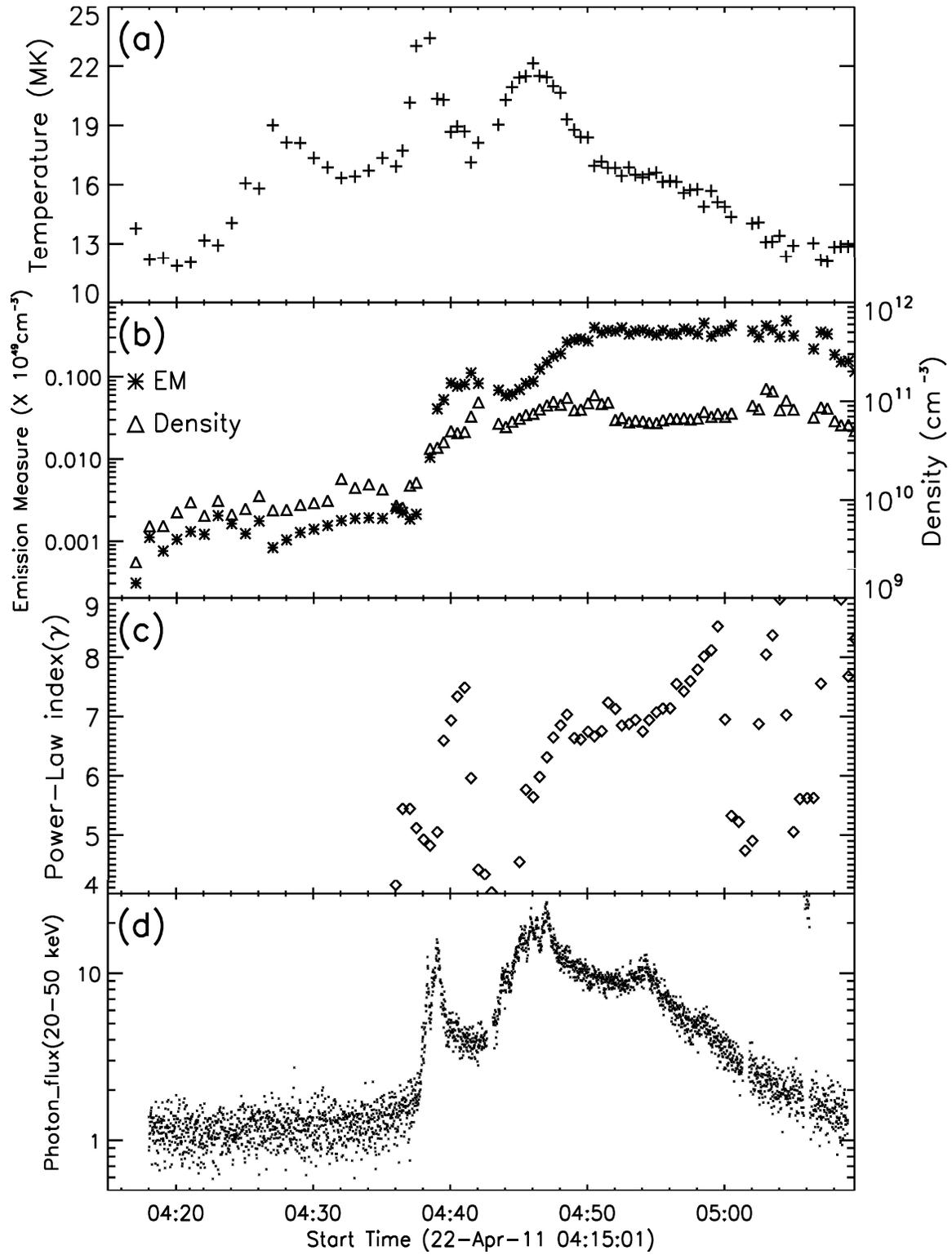}
  \caption{Panel (a): Temporal evolution of temperature (T), Panel (b): Left - EM; Right- density, Panel (c):  negative non-thermal power-law index ($\gamma$) and Panel (d): Intensity profile of emission in 20-50 keV from {\it RHESSI} observations.}
\end{figure*}

\subsection{Thermal and Non-thermal Energetics}
We estimate the thermal and non-thermal energies available during the precursor, impulsive and gradual phases of the flare evolution from the technique employed in \citet{2004JGRA..10910104E, 2005JGRA..11011103E} and \citet{2005A&A...435..743S}. For estimating thermal energy, we employ equation (1).
\begin{equation}
 E_{th}=3k_bT \sqrt{EM.V.f}
\end{equation}
where T and EM represent temperature and emission measure of flare plasma estimated from fitting the spectral observations in Section 3.2. V represents the volume of the flaring region and estimated by cylindrical approximation of the loops as discussed in Section 3.2. f represents the filling factor and assumed to be 1 which gives an estimation of upper limit of the thermal energy available in the flare. The thermal energy release rate is estimated to be varying in the range of 0.1- 4 $\times$ 10$^{29}$ ergs/s as shown in Fig. 7. The total thermal energy released during the flare is estimated to be $\sim$ 1.3 $\times$ 10$^{31}$ ergs out of which $\sim$2 per cent energy was released during the precursor phase and is estimated to be 2 $\times$ 10$^{29}$ ergs. \\

Further, we estimate non-thermal kinetic power considering the turnover model of electron flux as in equation (2)

\begin{equation}
P_{nth}= \int_{E_{min}}^{E_{to}} A_e E_{to}^{-\delta}dE + \int_{E_{to}}^{\infty} A_e E^{-\delta}dE
\end{equation}

Here, $\delta$ is the power-law index of the incident electron spectra and is estimated from the fitted power-law index $\gamma$ by $\delta$=$\gamma$+1. E$_{min}$ is the minimum energy for the electron flux and considered to be 6 keV for the current study. E$_{to}$ is the electron turn-over energy and estimated by the photon turn-over energy ($\epsilon_{to}$) and electron power-law index ($\delta$) employing the relation of fraction $\epsilon_{to}$/E$_{to}$ versus $\delta$ as derived in \citet{2005A&A...435..743S}. They estimated that the fraction $\epsilon_{to}$/E$_{to}$ varies between 0.4-0.6 for the values of $\delta$ ranging 3 to 8. With this relation, we estimate the interpolated value of aforesaid fraction for each $\delta$ to estimate E$_{to}$ corresponding to each $\epsilon_{to}$. Thus, E$_{to}$ is found to be varying in the range of 24-35 keV. Further, A$_{e}$ is obtained from the relation

\begin{center}
\begin{math}
A_\epsilon=\frac{A_e}{4 \pi D^2} \overline{z^2} \frac{\kappa _{BH}}{K} \frac{B(\delta -2, 1/2)}{(\delta -1 )(\delta -2)}
\end{math}
\end{center}

where A$\epsilon$ is the normalization index defined as

\begin{center}
\begin{math}
I_{thick}(\epsilon)=A_{\epsilon} \epsilon ^{-\gamma}
\end{math}
\end{center}

The non-thermal kinetic power (P$_{nth}$) for this flare is estimated which varies in the range of 0.01- 2 $\times$ 10$^{30}$ ergs/s and shown in Fig. 8. Further, total non-thermal energy released is estimated to be $\sim$ 4 $\times$ 10$^{31}$ ergs.

\begin{figure*}
  \centering
  \includegraphics[width=0.9\textwidth, height=0.4\textheight]{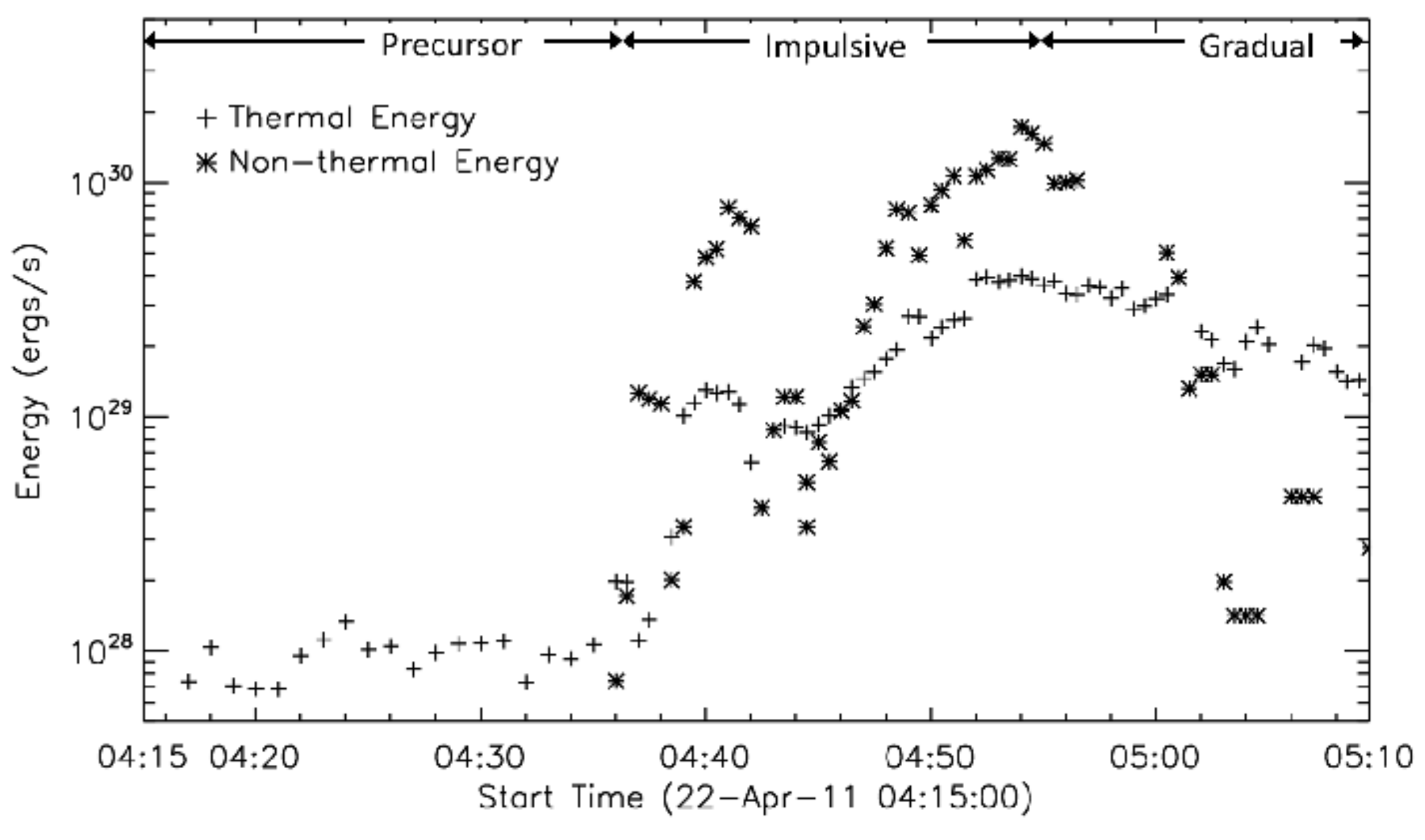}
  \caption{Temporal evolution of thermal and non-thermal energy release rate during various phases of the 2011 April 22 flare.}
\end{figure*}

We perform a comparative study of the energy released during various phases of emission. The total (thermal+non-thermal) energy ($E_{tot}$) released during the flare is estimated to be $\sim$ 5.3 $\times$ 10$^{31}$ ergs. During the precursor phase, only $\sim$1 per cent of E$_{tot}$ has been released which is originating from thermal processes. Further, during the impulsive phase, the rate of release of non-thermal energy is dominated by that of thermal energy and the fraction (Q$_f$) of non-thermal to thermal energy release rate (E$_{nth}$/E$_{th}$) varies in the range of 1 - 10 during this phase ({\it cf.} Fig. 7). However, after the onset of the gradual phase, the rate of release of non-thermal energy has been superseded by the thermal energy release rate and Q$_f$  reduces to $\sim$ 0.2, in agreement to \citet{2005A&A...438..325L}.

\subsection{Temperature and Emission Measure evolution}
We further study the spatial and temporal evolution of temperature (T) and emission measure (EM) employing multi-wavelength observations. We synthesize T and EM maps using the technique established by \citet{2013SoPh..283....5A}. The T and EM maps over the active region are synthesized using observations from the 6-channels of EUV viz. 94, 131, 171, 193, 211, 335 ~$ \rm \AA$ wavelengths. Firstly, the alignment of the co-temporal images obtained in 6 wavelengths with an accuracy of $\sim$1 pixel is performed by fitting the solar-limb detection algorithm\footnote{\url{http://www.lmsal.com/~aschwand/software/aia/aia_dem.html}}. The perfectly aligned images give a set of 6 EUV intensities at each pixel location over the AR which serves as input for forward-fitting modeled by Gaussian function to estimate emission measure-weighted temperature for a given pixel. This fitting is performed pixel-by-pixel to synthesize temperature and emission measure maps of the flaring region. We employ the aforesaid steps to the AIA observations obtained during 04:10:00 - 05:10:00 UT to synthesize the maps. The observations from {\it SDO/AIA} during the precursor phase were not saturated which allow us to restore the spatial evolution of the T and EM without any limitation. Although the observations have been saturated during the impulsive phase for some intermediate frames however the automatic exposure control onboard {\it SDO/AIA} usually alternated between short and long exposure times during the flare and therefore every second image is unsaturated and avoid problems in synthesizing the T and EM maps. We estimate peak temperature in the duration of synthesized map to be varying in the range of 5.7 - 10 MK. We overlay the contours of 80 per cent of the maximum intensity of 6-10, 10-14, 14-20 and 20-50 keV energy band on the T and EM maps as drawn by lines of green, blue, khaki and light orange colors, respectively as shown in Fig. 8. The top panel of Fig. 8 represents the spatial evolution of temperature at 04:05:03, 04:12:14, 04:19:26 and 04:38:14 UT, respectively. We further overlay the contour levels of 30 and 70 per cent of the maximum positive and negative magnetic-field observations from {\it SDO/HMI} over the T-map to track the polarity inversion line (PIL). The contours corresponding to positive and negative magnetic-fields are drawn with cyan and black lines respectively. It may be noted from the T-map synthesized at 04:05:03UT that the heating has started co-spatial to the polarity inversion line prior to the onset of SXR enhancement during the precursor phase. We further plot the EM-map in the middle and bottom panels of Fig. 8 during the precursor and main phases, respectively.

\begin{figure*}
  \centering
  \includegraphics[width=0.95\textwidth, height=0.46\textheight]{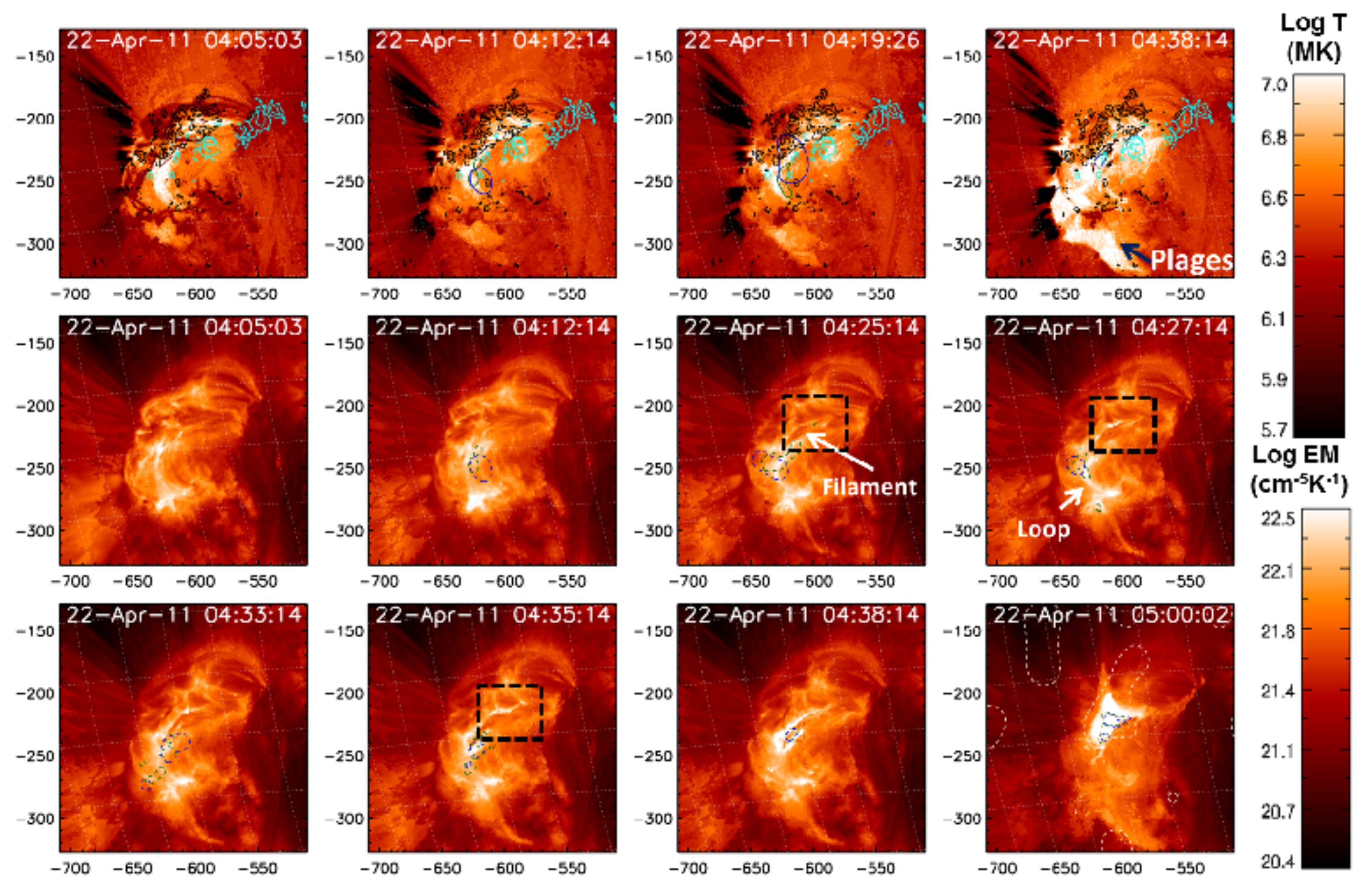}
  \caption{Temperature (top) and Emission measure (middle and bottom panel) maps. The temperature map is overlaid by HMI contours of leading and following polarities as represented by cyan and black colors respectively. The box in the middle and lower-panels represent the region of enhanced intensity prior to the eruption of the filament.}
\end{figure*}

The study of spatial evolution of the temperature with the time cadence of 12 s led us to various salient features: 1) The existence of heated plasma co-spatial to the polarity inversion line (PIL) during precursor phase; 2) Heated area has been found to be substantially increased during the impulsive phase in comparison to that during precursor phase; and 3) Plage brightening associated with the main phase with the peak temperature $\sim$10MK. Similarly from the synthesized EM maps we have been able to trace the filament location and morphological changes spatially associated with the T and EM.
Further, as previously mentioned, the loop-shape structure co-spatial to the SXR contours visible during the precursor phase can be observed in the middle panel of Fig. 8. We have noticed enhanced density in the vicinity of the north-west leg of the filament first appeared at 04:27:14 UT as highlighted in the dashed box in the middle panel of Fig. 8. In addition, we have also noted the filament activity in the form of slow-rise prior to the eruption, however almost parallel to the line-of-sight of the observations imposing the restriction on estimating the speed of eruption. Further, it is noteworthy that heated dense region has substantially enhanced at 04:35:14UT, $\sim$60 s before the onset of impulsive phase emission. This suggests that the small scale magnetic reconnection underneath the filament and associated heating have been going on prior to the accelerated eruption and the onset of impulsive phase of the flare.  We further study the morphological evolution of the filament in conjunction to the synthesized T and EM maps in the following section.

\subsection{$H \alpha$ emission and Filament morphology}
We use $H \alpha$ observations from ARIES/Nainital to study the role of the filament in the energy release during various phases of emission. Shown in Fig. 9 is a sequence of $\rm H \alpha$ images during various phases of emissions of the flare. The top panel shows a series of images representing the filament activity prior to the onset of the precursor phase at 04:01:15, 04:04:36, 04:05:52 and 04:10:54 UT respectively. We denote the filament legs into the southern as well as eastern direction as L1 and L2 respectively. It may be noted that the L1 started sequential movement towards east as marked in top panel of the figure during 04:05:52 - 04:10:54 UT. Further, L1 possibly erupted prior to the onset of emission in the precursor phase and disappeared as may be seen in the frame at 04:15:05 UT in the middle panel. In addition to the partial eruption of the filament, we have also noticed the filament activity in the form of sigmoidal shape and a clear appearance of other leg during 04:29:48 - 04:36:00 UT as represented by L3 in the middle panel. The bottom panel of the Fig. 9 represents the time sequence of $H \alpha$ filtergrams during main phase of the flare at 04:36:04, 04:37:45, 04:40:07 and 04:56:02 UT, respectively. We note that the enhanced emission visible at 04:37:45 UT which expanded later during the gradual phase.

\begin{figure*}
  \centering
  \includegraphics[width=0.95\textwidth, height=0.5\textheight]{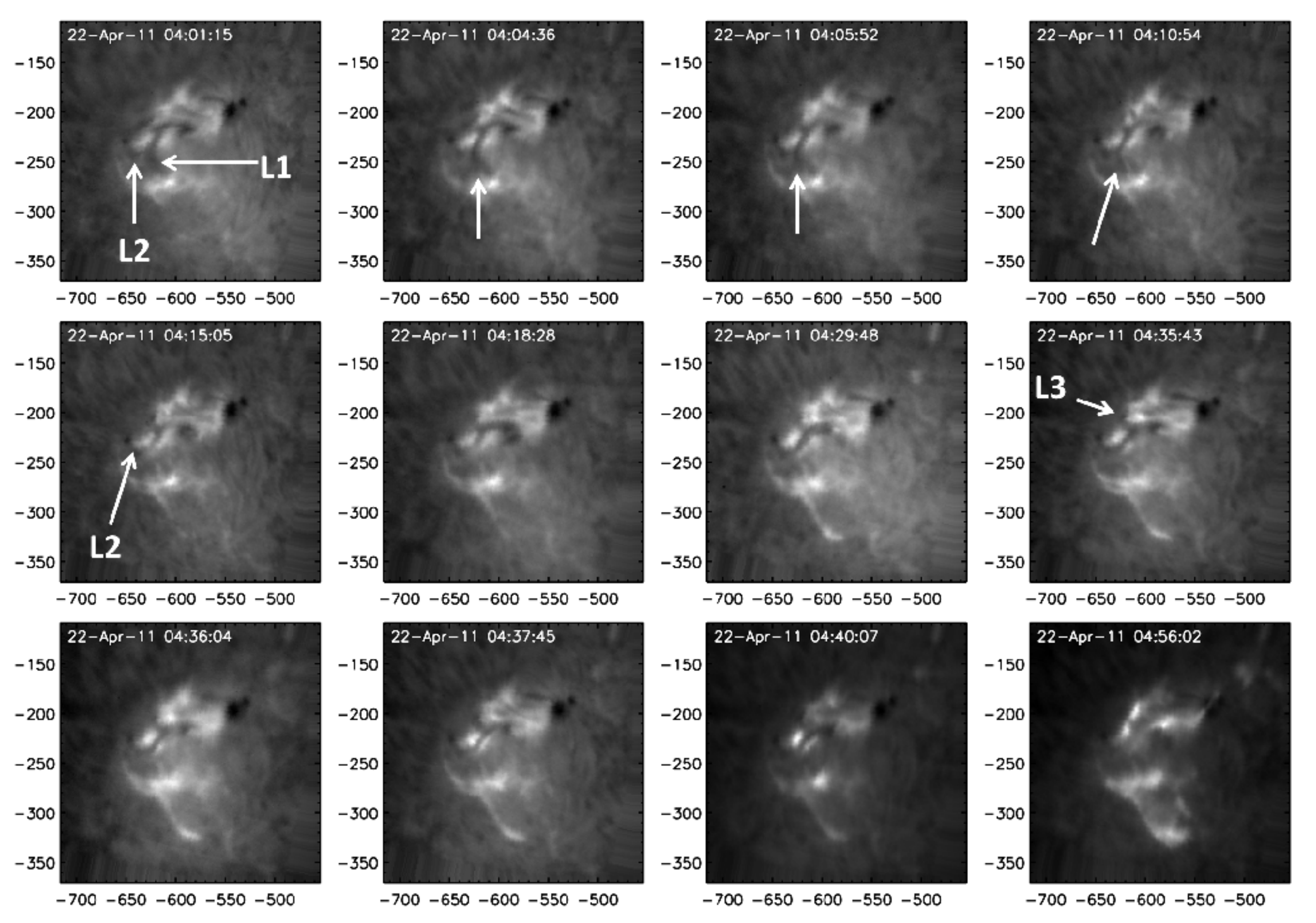}
  \caption{Top panel - Sequence of images representing the filament activity prior to the onset of precursor phase at 04:01:15, 04:04:36, 04:05:52 and 04:10:54 UT, respectively. Middle and bottom panels represent the evolution of filament during the precursor and main phases, respectively.}
\end{figure*}

The study of the spatial evolution of the filament reveals the role of the partial filament eruption in triggering the precursor phase emission and later the whole filament system has been destabilized prior to the onset of the main phase emission. In order to probe the cause of destabilization of the filament we investigate the morphological evolution of filament in conjunction to the T and EM maps. Fig. 10 shows the time series of $H \alpha$ observations from ARIES/Nainital overlaid by the contours of log T= 6.8 and log EM= 22.5 drawn by red and green colors, respectively. We may note the heating of the filament from the top panel of Fig. 10 which represents the evolution of T and EM with H-alpha intensity during precursor phase at 04:10:54, 04:18:28 and 04:29:48 UT respectively. Further, the heating remains co-spatially associated with the filament during the whole precursor phase which, however, followed an expansion during the main phase after the destabilization of the filament. In addition, contours representing EM evolution in the top panel clearly depict the loop-shaped structure during the precursor phase. The evolution of loop emission during precursor phase has also clearly been visible in 131~$ \rm \AA$ emission.
\begin{figure*}
  \centering
  \includegraphics[width=0.9\textwidth, height=0.4\textheight]{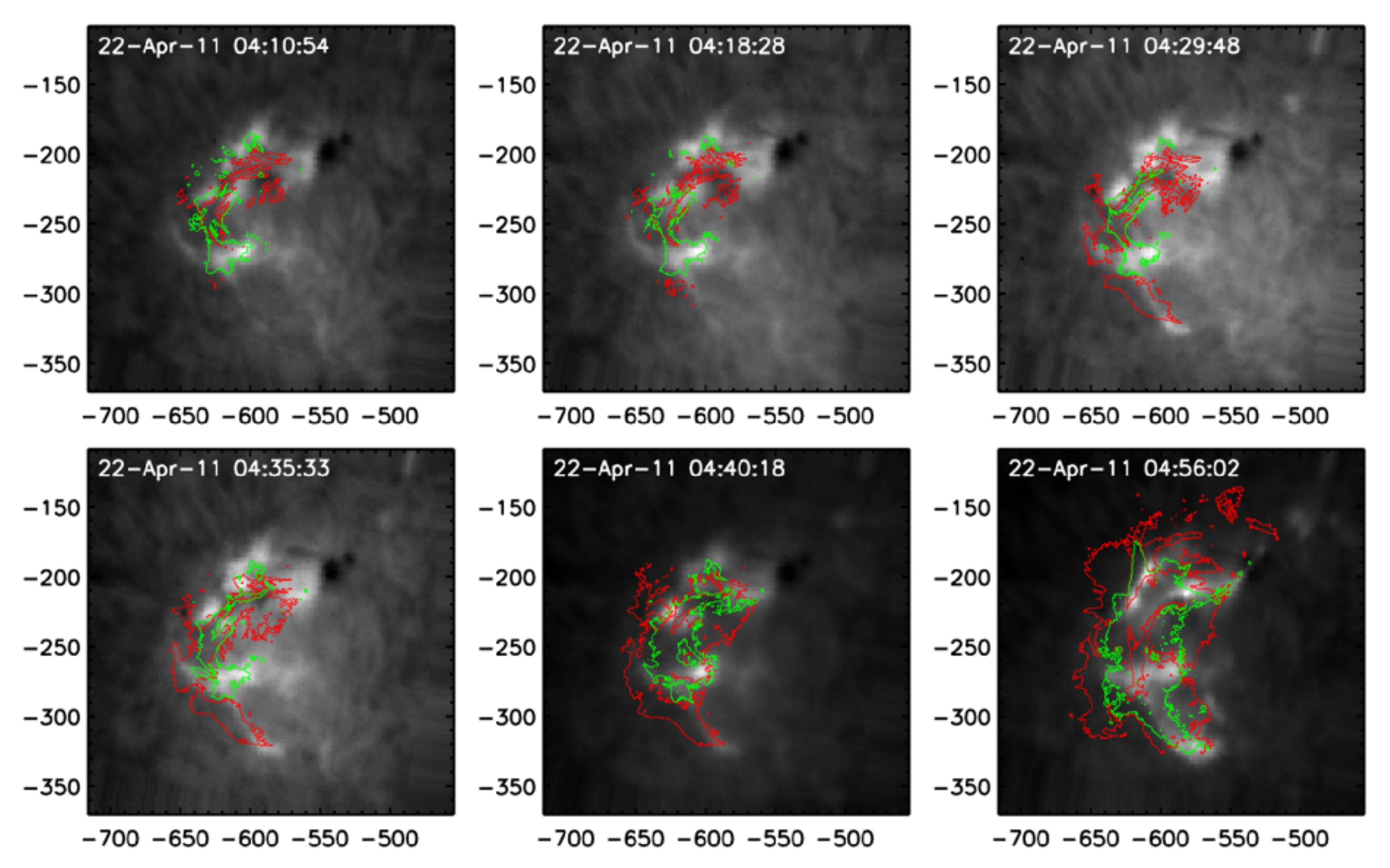}
  \caption{Sequence of $H \alpha$ images (ARIES/Nainital) overlaid by the contours of log T=6.8 and log EM=22.5 drawn by red and green lines, respectively.}
\end{figure*}

\section{Summary and Conclusions}

We have investigated the spatial, spectral and temporal evolution of multi-wavelength emission in various phases of the M1.8 flare occurred on 2011 April 22 in NOAA AR 11195. In this section we combine the results obtained from multi-wavelength analysis and interpretations to propose a unified scheme of energy release during the precursor and main phases. We present the key results of the study and associated implications as following:

\begin{figure*}
  \centering
  \includegraphics[width=0.9\textwidth, height=0.21\textheight]{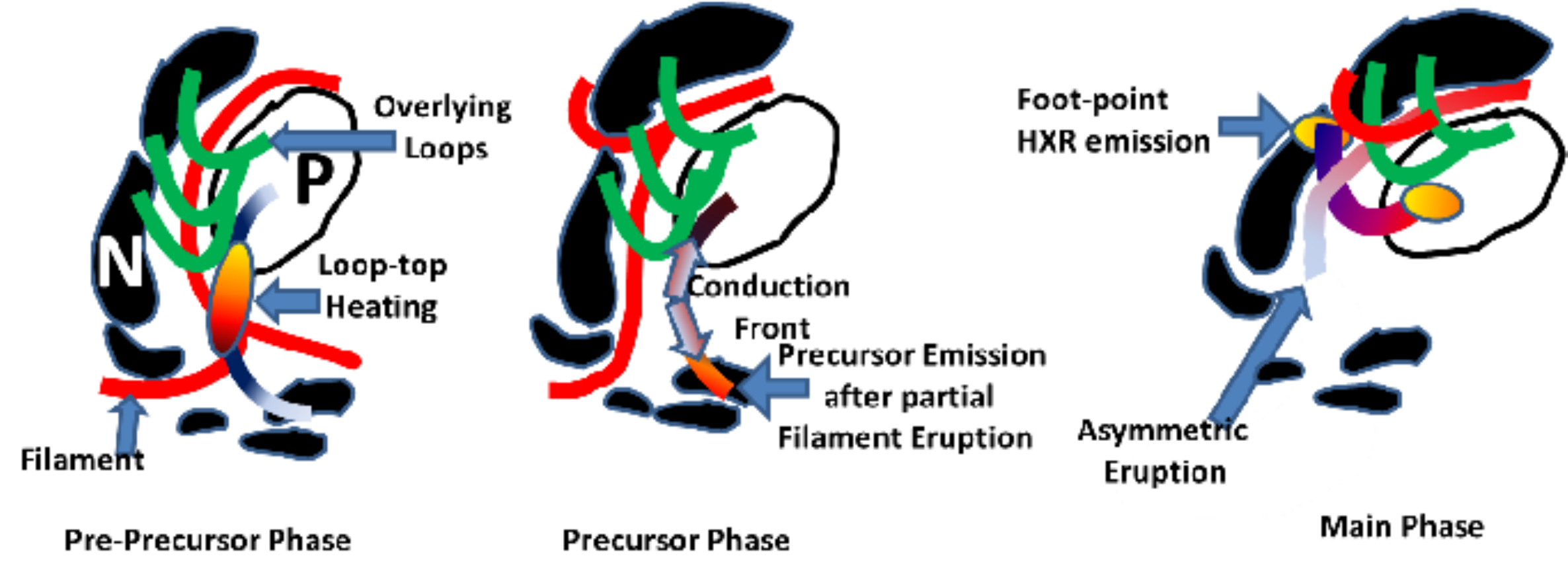}
  \caption{Schematics of the physical processes during the M1.8 flare from AR11195. The filament is shown by the thick red line. The white blob (P) represents the positive polarity and the black scattered blobs (N) represent the negative polarities. The curves filled with gradient colors represent the emitting loops while the green curves represent the overlying loops associated with the AR and the constituents of the main phase emission.}
\end{figure*}

\begin{enumerate}

\item The study of spatial evolution of multi-wavelength emission during the precursor and main phases has been performed. We note a brightened loop-top in 131~$ \rm \AA$ observations prior to the onset of the precursor phase and later uniform loop emission during the precursor phase whereas no foot-point emission was observed during this phase. Therefore, the appearance of coronal source heating prior to the precursor emission suggests the possibility of conduction-driven chromospheric heating during this phase. In contrast, the foot-point HXR sources in addition to the SXR loops have been pronounced during the main phase emission which can be explained by the standard model of energy release in solar flares.

\item The spectral fitting has been performed on the {\it RHESSI} observations over the flare duration to study the evolution of flare plasma parameters viz. temperature, emission measure and power-law index. The estimated temperature during the precursor phase attained its peak value of $\sim$ 19 MK in comparison to that in the impulsive phase of $\sim$23 MK. We have also estimated the T and EM values from {\it SDO/AIA} observations. We note that the T values estimated from {\it RHESSI} are twice as high as those from SDO/AIA differential emission measure peak temperature. This difference in the temperature values may be attributed to the fact that current investigation employs isothermal photon function for estimating the flare plasma temperature from {\it RHESSI} observations which leads to the overestimated values of T as demonstrated in the recent studies by Ryan D. F. et al. (Private Communication) and Aschwanden M. J. \& Shimiju, T. (Private Communication).

\item The spectral fitting has also revealed that the rate of increase of EM during the precursor phase has been slow in comparison to that during the main phase. On the other hand, the estimated density ($n_e$) has remained almost constant at $\sim$ 10$^{10}$ cm$^{-3}$ during the precursor phase however increased drastically to $\sim$ 10$^{11}$ cm$^{-3}$ after the commencement of the impulsive phase. The temporal evolution of T and $n_e$ suggests a relatively slow heating rate of the chromosphere most likely to be due to gentle evaporation by conduction front, which has also been suggested by \citet{2009A&A...498..891B}.

\item The temporal evolution of density during the precursor and gradual phases does not show rapid changes in contrast to the evolution during the impulsive phase. This suggests that chromospheric evaporation is gradual in both phases. Earlier, \citet{2009ApJ...701.1209B} also explored the scenario of gentle evaporation, however, only during the gradual phase.

\item The study the temporal evolution of relative intensity of several $H \alpha$ brightening within the flaring region and its correlation with the X-ray emission is carried out. The onset of the precursor phase emission in $H \alpha$ at 04:18:24 UT is unambiguous. Further, the first peak in the $H \alpha$ emission $P_{h1}$ is delayed by $\approx$ 134 s relative to that in 6-20 keV emission denoted by $P_{x1}$. Theoretically, the delay of the peak in $H \alpha$ intensity profile with respect to the corresponding peak in the X-ray intensity profile represents the response time of the chromosphere which is heated either by the accelerated electrons or by the conduction front produced during magnetic reconnection. Considering the fact that this heating is produced by the non-thermal particles' interaction with the chromosphere, the delay is short ($\sim$10 s) which, on the other hand, turns out to be $\sim$20 s in case of conduction-driven chromospheric evaporation \citep{2011A&A...535A.123R}. The extra-ordinary long delay of 134 s may point towards the phenomena of slow response of the chromosphere undergoing through gentle evaporation originated by conduction-front during the precursor phase. We estimate the half loop-length ($L_{loop}$) from 131 ~$ \rm \AA$ image during the precursor phase to be $\sim$ 30,000 km which, therefore, suggests the conduction front movement speed ($v_{cond}=L_{loop}/\Delta t$) to be $\sim$ 450 km/s. This estimated velocity estimation may provide direct evidence to the numerically estimated speed of the conduction front ranging between 450 and 600 km/s by \citet{2010AAS...21632004G}. Further, during the impulsive phase, $P_{h2}$ has been delayed by $\sim$10 s with respect to $P_{x2}$ which confirms the presence of non-thermal interaction with the chromosphere during this phase. The impulsive phase has shown two peaks at 04:39:16 UT ($P_{x2}$) and at 04:47:10 UT in HXR emission. Further, we may note a plateau region of HXR emission during 04:47:30 - 04:54:30 UT, which suggests the continuation of reconnection.Further, as the HXR emission is leveled off and the gradual phase onsets at 04:54:20 UT, third and the highest peak in $H \alpha$ emission, denoted as Ph3 in Fig. 1, at 04:54:30 UT appears. This temporal comparison suggests that chromospheric heating due to non-thermal electrons started from the onset of impulsive phase and continued until the plateau duration ended.

\item To understand the source of the energy made available during the precursor phase, we have undertaken the study of spatial evolution of T and EM in conjunction with the $H \alpha$ filtergrams which has revealed that the filament was heating-up during the precursor phase. Further, we have noticed the eruption of one leg of the filament L1 prior to the onset of precursor phase. We also find slow movement in the filament leading to a sigmoid shape as well as increase in the $H \alpha$ intensity during the precursor phase. \item This suggest that after the partial filament eruption leading to the precursor phase emission, the magnetic-field restructuring has occurred which led to the more complex magnetic connectivity as depicted by sigmoid shaped filament. Thus the filament underwent heating and slow-rise throughout the precursor phase caused by magnetic reconnection underneath the filament \citep{2009SoPh..256...57L} and the main phase emission has been triggered by the eruption of the unstable filament. We also note the enhancement of EM in the vicinity of the north-east leg of the filament before the onset of filament eruption which assist the scenario presented here for the eruption of the filament.
\end{enumerate}

Based on the study and interpretations, we propose a unified scheme of energy release during the precursor and main phase emission in the flare event of investigation. Figure 11 shows the schematics of the phenomenological evolution during the M1.8 flare of current study. The filament is shown by the thick red line. The white blob (P) represents the positive polarity and the black scattered blobs (N) represent the negative polarities. The curves filled with gradient colors represent the emitting loops while the uniform green curves represent the overlying loops associated with the AR and the constituents of the main phase emission. In this unified scheme, the precursor phase emission originates via the conduction front triggered by the partial filament eruption. Next, the heated leftover S-shaped filament has undergone slow rise and heating due to magnetic reconnection and finally erupted to produce emission during the impulsive and gradual phases.

\section*{Acknowledgments}
The authors thank IUSSTF/JC-Solar Eruptive Phenomena/99-2010/2011 - 2012 project on ``Multi-wavelength Study of Solar Eruptive Phenomena and their Interplanetary Responses'' for its support to this study. We acknowledge the free data usage policy of the {\it SDO/AIA} and {\it SDO/HMI} as well as {\it STEREO}, {\it SOXS} and {\it RHESSI} missions. AKA, RJ and PG acknowledge the support of Dept. of Space (Govt. of India). AKA and RJ acknowledge the guidance provided by Prof. Brain Dennis during the {\it RHESSI} data analysis. The numerical computations have been performed on the three TFLOPs cluster at PRL. This work is also a part of the project carried out under the Climate and Weather of the Sun-Earth System (CAWSES)-India Program supported by the Indian Space Research Organization (ISRO). The authors also acknowledge the support of the anonymous referee in the form of descriptive comments which helped in improving the manuscript.

%\bibliography{bib.bbl}
%\bibliographystyle{abbrvat}
%\input{bib.bbl}

\label{lastpage}

\end{document}